\documentclass[prl,twocolumn,showpacs,superscriptaddress,groupedaddress]{revtex4-1}
\usepackage{amsfonts}
\usepackage{amsmath}
\usepackage{amssymb}
\usepackage{graphicx}

\input{tcilatex}
\begin{document}

\title{Quantum crystals in a trapped Rydberg-dressed Bose-Einstein condensate%
}
\author{C.-H. Hsueh$^{1}$, T.-C. Lin$^{2}$, T.-L. Horng$^{3}$, and W. C. Wu$%
^{1}$}
\date{\today}

\begin{abstract}
Spontaneously crystalline ground states, called quantum crystals, of a
trapped Rydberg-dressed Bose-Einstein condensate are numerically
investigated. As a result described by a mean-field order parameter, such
states simultaneously possess crystalline and superfluid properties. A
hexagonal droplet lattice is observed in a quasi-two-dimensional system when
dressing interaction is sufficiently strong. Onset of these states is
characterized by a drastic drop of the non-classical rotational inertia
proposed by Leggett [Phys. Rev. Lett. \textbf{25}, 1543 (1970)]. In
addition, an AB stacking bilayer lattice can also be attained. Due to an
anisotropic interaction possibly induced by an external electric field,
transition from a hexagonal to a nearly square droplet lattice is also
observed.
\end{abstract}

\maketitle

\affiliation{$^1$Department of Physics, National Taiwan Normal University, Taipei 11677, Taiwan\\
$^2$Department of Mathematics, National Taiwan University, Taipei, 10617, Taiwan\\
$^3$Department of Applied Mathematics, Feng Chia University, Taichung 40724, Taiwan}

Superfluidity which implies a long-range phase coherence is a crucial
property at low temperatures of many quantum liquids or gases such as liquid
Helium or Bose-Einstein condensates (BECs), whereas crystallization implies
a long-range configurational order. Superfluidity and crystallization are
generally two conflicting properties. Penrose and Onsager \cite%
{PhysRev.104.576} were the first to consider a BEC in a solid and concluded
that such a \textit{supersolid} state simultaneously possessing crystalline
and superfluid properties was impossible. Since then, this question has been
revisited by a number of authors \cite{Lifshitz,Chester,Legget} and has been
a matter of large speculation for the last forty years. Recently, the
observation of a supersolid phase in $^{4}$He systems \cite{supersolid in He}
revitalized this fundamental interest.

An alternative and excellent candidate to study supersolidity is in atomic
BECs which provide a clean and experimentally controllable system. The
crystal structure in solid helium can be replaced by the modulated density
in BEC. Density modulated BECs are already formed by the imposition of an
external potential, creating the so-called optical lattices. In these
systems, by varying the properties of the optical lattice, the condensate
was shown to exhibit a Mott insulator/superfluid phase transition \cite%
{OL1,OL2,OL3,OL4}. More recently, it has been shown that supersolidity might
be present for Rydberg atoms in the dipole/van der Waals (vdW) blockade
regime \cite{InTrap,hexagonal,heaviside,3D}. Cinti \textit{et al}.\cite%
{hexagonal} considered a dipole-dipole interaction softening at short
distance, allowing for a ground-state computation that happens to display
the properties of supersolidity. It proved that a quantum system of
interacting particles can exhibit both crystalline structure and
superfluidity property. Similar results were obtained by Saccani \emph{et al}%
.\cite{heaviside} by using a Heaviside-function interaction. Based on a
mean-field treatment, Henkel \textit{et al}.\cite{3D} proposed that a BEC of
particles interacting through an isotropically repulsive vdW interaction
with a softened core might support a density modulation. They found that the
Fourier transform of such interaction has a partial attraction in momentum
space, which gives rise to a transition from a homogeneous BEC to a
supersolid phase because of the roton instability (see also Refs.~\cite%
{Rica,Lin}).

Based on the Gross-Pitaevskii (GP) treatment, this Letter aims to study the
ground states of a trapped Rydberg-dressed BEC. Comparing to other GP work
that did not consider the effect of trapping \cite{3D}, we exactly solve the
nonlocal GPE with a trap. In particular, we focus on the quasi-2D geometry
and the strong interacting regime that lead to various fascinating
ground-state structures of Rydberg-dressed BEC. 
It will be shown that in the supersolid phase, periodic structures of
Rydberg-dressed BEC can undergo a transition from concentric rings to a
lattice (or crystalline) if the interaction is above some critical value.
The lattices are formed in terms of crystalline superfluid droplets, called
quantum crystals, whose onset is characterized by a drastic drop of the
non-classical rotational inertia fraction (NCRIF) \cite{Legget}. The
crystalline structure appears to be a hexagonal lattice in a
quasi-two-dimensional (quasi-2D) geometry which can turn into a nearly
square lattice if interaction acquires an anisotropic component in the
presence of an external electric field (Stark effect). Moreover, multilayer
crystal structure such as an AB stacking bilayer is also obtained when the
frozen axis is relaxed or particle number is increased. Crystalline in the
case of a quasi-one-dimensional (quasi-1D) geometry will also be presented.

We start from a nonlocal Gross-Pitaevskii equation (GPE) 
\begin{eqnarray}
i\hbar \partial _{t}\Psi \left( \mathbf{r},t\right) &=&\left[ -\frac{\hbar
^{2}\nabla ^{2}}{2M}+V_{\mathrm{ext}}\left( \mathbf{r}\right) +g\left\vert
\Psi \left( \mathbf{r},t\right) \right\vert ^{2}\right.  \notag \\
&+&\left. \int U\left( \mathbf{r}-\mathbf{r}^{\prime }\right) \left\vert
\Psi \left( \mathbf{r}^{\prime },t\right) \right\vert ^{2}d\mathbf{r}%
^{\prime }\right] \Psi \left( \mathbf{r},t\right) ,  \label{nonlocal GPE}
\end{eqnarray}%
where $U(\mathbf{r}-\mathbf{r}^{\prime })=\tilde{C}_{6}/(R_{c}^{6}+ \vert 
\mathbf{r}-\mathbf{r}^{\prime }\vert ^{6})$ is an isotropically repulsive
vdW interaction between Rydberg-dressed ground-state atoms \cite%
{InterRdAtoms} with $\tilde{C}_{6}$ and $R_{c}$ the effective coupling
constant and blockade radius respectively (we will return to the interaction
later). Here $\Psi$, normalized as $N=\int\vert \Psi\vert ^{2}d\mathbf{r}$ ($%
N$ is the atom number), denotes the condensate wave function of dressed
atoms. $M$ denotes the atomic mass and $g=4\pi \hbar ^{2}a/M$ describes the
strength of local interaction due to $s$-wave scattering with scattering
length $a$. In the cylindrical coordinates, ($\rho ,\phi ,z$), the harmonic
trapping potential $V_{\mathrm{ext}}(\mathbf{r})=(M\omega _{\perp
}^{2}/2)(\rho ^{2}+\lambda ^{2}z^{2})$ with radius frequency $\omega _{\perp
}$ and aspect ratio $\lambda $ is included for simulating real experiments.

By introducing useful length ($R_{c}$) and time ($\tau \equiv
R_{c}^{2}M/\hbar $) scales, Eq.~(\ref{nonlocal GPE}) can be rewritten as%
\begin{eqnarray}
&&i\partial _{t}\psi \left( \mathbf{r},t\right) =\left[ -\frac{\nabla ^{2}}{2%
}+\frac{\omega ^{2}\left( \rho ^{2}+\lambda ^{2}z^{2}\right) }{2}\right. 
\notag \\
&&~~~~+\gamma \left. \left\vert \psi \left( \mathbf{r},t\right) \right\vert
^{2}+\alpha \int \frac{\left\vert \psi \left( \mathbf{r}^{\prime },t\right)
\right\vert ^{2}d\mathbf{r}^{\prime }}{1+\left\vert \mathbf{r}-\mathbf{r}%
^{\prime }\right\vert ^{6}}\right] \psi \left( \mathbf{r},t\right) ,
\label{dimesionless}
\end{eqnarray}%
where we have redefined the normalized wave function $\psi \equiv \sqrt{%
R_{c}^{3}/N}\Psi $, the strength of the radius potential $\omega \equiv
\omega _{\perp }\tau $, and the interaction constants $\gamma \equiv 4\pi
Na/R_{c}$ and $\alpha \equiv MN\tilde{C}_{6}/(\hbar ^{2}R_{c}^{4})$. To
obtain ground-state wave functions, we computed the governing Eq.~(\ref%
{dimesionless}) with imaginary time propagation till the convergence of the
normalized wave function with error less than $10^{-6}$. Moreover, we have
used the method of lines with spatial discretization by the Fourier
pseudospectral method. The time integration in Eq.~(\ref{dimesionless}) is
done by the adaptive Runge-Kutta method of order 2 and 3 (RK23), which is
more time efficient due to an adjustable time step.

\begin{figure}[tbp]
\begin{center}
\includegraphics[height=3.605in,width=3.1401in]{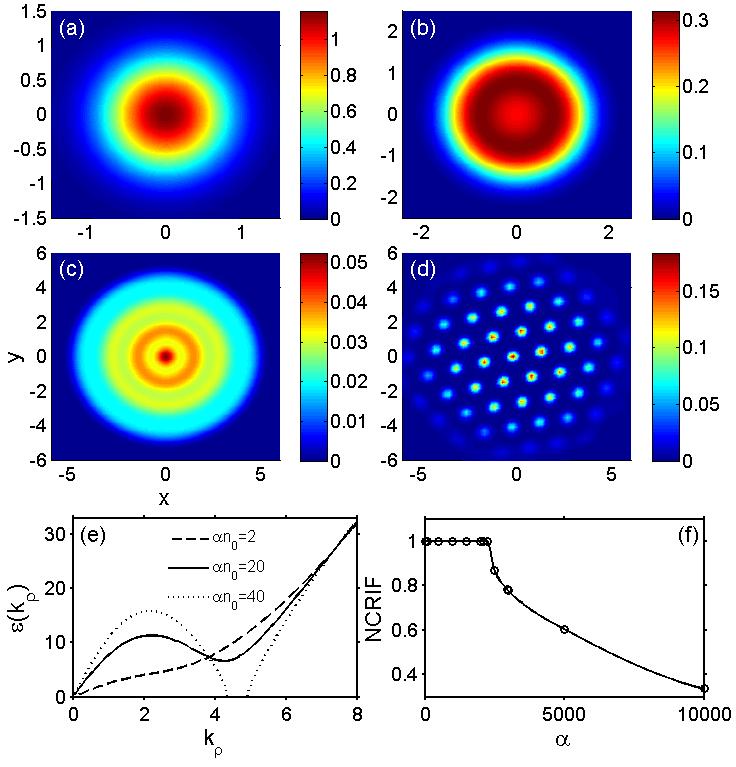}
\end{center}
\par
\vspace{-0.8cm}
\caption{(Color online) Density modulations of the quasi-2D Rydberg-dressed
condensate with $\protect\omega =3$, $\protect\lambda =8$, $\protect\gamma =0
$, and $\protect\alpha =10$ (a), $50$ (b), $2000$ (c), and $5000$ (d). (e)
plots the dispersion relation $\protect\epsilon (k_{\protect\rho })$ [Eq.~(%
\protect\ref{dispersion})] for various interaction coupling $\protect\alpha $
in a uniform 2D limit. Roton instability occurs when $\protect\alpha n_{0}
\agt37$. (f) shows the calculated NCRIF [Eq.~(\protect\ref{NCRIF})] with
rotation velocity $\protect\omega _{0}=0.01$, signifying crystallization at $%
\protect\alpha \sim 2300$.}
\label{fig1}
\end{figure}

For the Rydberg-dressed BEC trapped in a harmonic potential, 
one can define useful characteristic lengths, $a_{\perp }\equiv \sqrt{\hbar
/( M\omega _{\perp }) }$ and $a_{\Vert }\equiv a_{\perp }/\lambda $,
corresponding to radial and axial potential, respectively. The spectrum and
the onset of instability are tunable by varying the particle number or the
confining potential. By varying the two ratios $R_{c}/a_{\perp }=\sqrt{%
\omega }$ and $R_{c}/a_{\parallel }=\sqrt{\lambda \omega }$, one can
effectively have quasi-1D ($\sqrt{\omega }\gg 1$ and $\sqrt{\lambda \omega }%
\approx 1$) or quasi-2D ($\sqrt{\omega }\approx 1$ and $\sqrt{\lambda \omega 
}\gg 1$) limits.

A quasi-2D system with $\omega =3$ and $\lambda =8$ is first studied. Fig.~%
\ref{fig1} shows the condensate density profile varying with the strength of
dressing-induced interaction $\alpha $. When $\alpha $ is small, the system
displays superfluidity, and the ground-state density profile exhibits a
central peak [see Fig.~\ref{fig1}(a)]. As $\alpha $ increases, the central
density is too high to be stable and thus starts to modulate. Fig.~\ref{fig1}%
(b) shows a cratered condensate due to the central instability. As $\alpha$
increases further, owing to the roton instability occurring in the modulated
density (see later), condensate wrinkles violently and forms a ring
structure [see Fig.~\ref{fig1}(c)]. When $\alpha $ is increased above a
critical value $\sim 2300$, condensate eventually forms a droplet lattice.
Fig.~\ref{fig1}(d) shows the quasi-2D Rydberg-dressed BEC forming a
hexagonal droplet lattice.

It is important to check whether the parameter regime discussed above is
actually experimentally accessible. The vdW interaction $U(\mathbf{r}-%
\mathbf{r}^{\prime})$ can be generated from the off-resonant dressing of
ground-state atoms with high-lying Rydberg states. Considering, for example,
ground-state $^{87}$Rb atoms coupled to excited Rydberg $nS$ state $^{87}$Rb
atoms with $n=60$ via a Rabi frequency $\Omega$ and a red laser detunning $%
\Delta <0$, it will admix a small fraction $\nu =(\Omega /2\Delta )^{2}$ of
Rydberg character into the ground-state atoms for weak dressing ($\nu \ll 1$%
). For two far-distant atoms, it leads to an effective interaction $\tilde{C}%
_{6}/r^{6}$ ($\tilde{C}_{6}=\nu ^{2}C_{6}$), arising from the strong vdW
interaction $C_{6}/r^{6}$ ($C_{6}=9.7\times 10^{20}$ a.u.) \cite{Singer}. At
shorter distances, the two atoms enter the vdW blockade regime \cite{vdW} to
which the effective interaction saturates. Altogether, as given earlier, the
effective potential between Rydberg-dressed ground-state atoms is $U(\mathbf{%
r}-\mathbf{r}^{\prime})=\tilde{C}_{6}/(R_{c}^{6}+|\mathbf{r}-\mathbf{r}%
^{\prime }|^{6})$ with the blockade radius $R_{c}=(C_{6}/2\hbar |\Delta
|)^{1/6}$ \cite{3D}. As shown, the above off-resonant scheme produces only
repulsive nonlinearities for alkaline atoms \cite{Rydberg}. Using typical
value of Rabi frequency $\Omega=580$KHz and a red detunning $|\Delta |=50$%
MHz, blockade radius $R_{c}=4.5$ $\mu$m. Moreover, the effective lifetime of
dressed atoms, $1/\gamma _{\mathrm{eff}}\equiv 1/(\nu \gamma _{\mathrm{r}})$%
, is as large as several seconds with Rydberg state decaying rate $\gamma_{%
\mathrm{r}}\sim 10$ms$^{-1}$ \cite{InterRdAtoms}. To achieve the largest
coupling constant that we are considering, $\alpha =5000$, based on the
above parameters one needs the total atom number $N=1.5\times 10^{6}$ which
corresponds to $N_{\mathrm{r}}=\nu N=52$ for the number of excited Rydberg
atoms. By counting the number of droplets $\sim 50$ in Fig.~\ref{fig1}(d),
an average of one excited Rydberg atom together with $3\times 10^{4}$
ground-state atoms is within a single droplet. This justifies the validity
of the GP treatment with the two-body dressing interaction $U(\mathbf{r}-%
\mathbf{r}^{\prime})$ \cite{critical}.

\begin{figure}[tbp]
\begin{center}
\includegraphics[height=1.2842in,width=3.1393in]{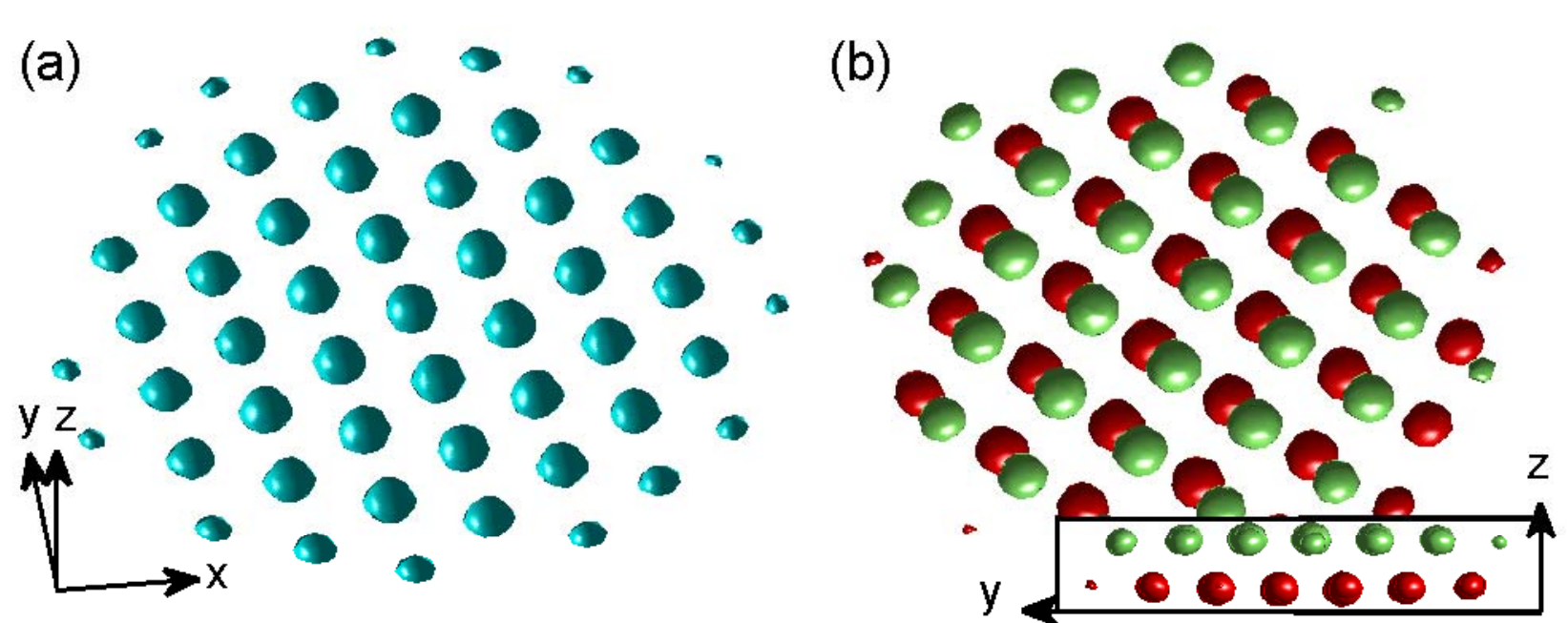}
\end{center}
\par
\vspace{-0.5cm}
\caption{(Color online) Comparison of a monolayer (a) and a bilayer (b)
lattice structures. (a) is the isosurface (isovalue $0.03$) of the density
with $\protect\omega =3$, $\protect\lambda =8$, $\protect\gamma =0$, and $%
\protect\alpha =5000$, while (b) is the isosurface (isovalue $0.015$) of the
density with $\protect\omega =3$, $\protect\lambda =5$, $\protect\gamma =0$,
and $\protect\alpha =10000$. Inset in (b) shows the bilayer being an AB
stack.}
\label{fig2}
\end{figure}

A qualitative understanding of forming the ring-like structures [Fig.~\ref%
{fig1}(c)] in a quasi-2D system is given in the uniform 2D limit.
Considering that the system is strongly confined in $z$-direction by a
harmonic potential, $(\omega \lambda )^{2}z^{2}/2$ but free moving in the $%
xy $ plane, superfluid BEC density can be approximated by $n(\mathbf{r}%
)=n_{0}\varphi _{g}^{2}(z)=n_{0}\sqrt{\omega \lambda /\pi }\exp (-\omega
\lambda z^{2})$. The 2D excitation spectrum calculated from the
corresponding Bogoliubov-de Gennes equations is thus 
\begin{equation}
\varepsilon \left( k_{\rho }\right) =\sqrt{\frac{k_{\rho }^{2}}{2}\left[ 
\frac{k_{\rho }^{2}}{2}+2\alpha n_{0}\tilde{U}_{\mathrm{2D}}(k_{\rho })%
\right] },  \label{dispersion}
\end{equation}%
where $k_{\rho }=|\mathbf{k}_{\rho }|$ (in units of $1/R_{c}$) is momentum
component in the $xy$ plane and $\tilde{U}_{\mathrm{2D}}(k_{\rho })= \int 
\tilde{U}(k)\tilde{n}_{g}(k_{z})\tilde{n}_{g}(-k_{z})dk_{z}/2\pi $ with $%
\tilde{n}_{g}(k_{z})=\exp [-k_{z}^{2}/(4\omega \lambda )]$ and $\tilde{U}%
(k)=(2\pi ^{2}/3)(e^{-k/2}/k)[e^{-k/2}-2\sin (\pi /6-\sqrt{3/4}k)]$ ($k=%
\sqrt{k_{\rho }^{2}+k_{z}^{2}}$) being Fourier transforms of the product of $%
\varphi _{g}^{2}$ and the scaled interaction $1/(1+r^{6})$ in Eq.~(\ref%
{dimesionless}) \cite{Uk2D}. When deriving Eq.~(\ref{dispersion}) and
throughout this Letter, contact interaction $\gamma$ is set to zero.
Including $\gamma $ will not affect the behaviors of roton instability if $%
\alpha $ is sufficiently large, nor will affect the phonon behavior if $%
\alpha $ is small to which leading term of the expansion in $\tilde{U}_{%
\mathrm{2D}}(k_{\rho })$ is a positive constant, i.e., nonzero $\gamma $
only modifies the sound velocity. $\varepsilon (k_{\rho })$ has
asymptotically a phonon and a free-particle character at small and large $%
k_{\rho }$, respectively. However, with $\tilde{U}_{\mathrm{2D}}(k_{\rho })$
having a negative minimum at some finite momentum, $\varepsilon (k_{\rho })$
drops near that particular momentum, and eventually becomes imaginary when
increasing the strength $\alpha $ [see Fig.~\ref{fig1}(e)]. It is estimated
that when $\alpha n_{0}\agt37$, the assumed uniform superfluid state is
unstable towards formation of nonuniform (periodic ring-like) structures. 

To characterize the transition from concentric rings to a crystalline
hexagonal lattice, we study the non-classical rotational inertia fraction
(NCRIF), defined by $(I_{0}-I)/I_{0}$. Here $I$ is moment of inertia of the
superfluid system under study and $I_{0}$ is its corresponding classical
value \cite{supersolid in He}. As proposed by Leggett \cite{Legget}, NCRIF
of the superfluid system can be calculated under a small rotation. In the
rotating frame, free energy of a rotating BEC with rotation velocity $\omega
_{0}$ about $z$ axis is 
\begin{eqnarray}
F(\omega_{0})&=&F_{0}-\omega _{0}\langle \psi ,L_{z}\psi \rangle  \notag \\
&=&\int \left[ \frac{\left\vert \left( \nabla -i\omega _{0}\mathbf{e}%
_{z}\times \mathbf{r}\right)\psi(\mathbf{r})\right\vert ^{2}}{2}+\frac{%
\omega ^{2}\left( \rho ^{2}+\lambda ^{2}z^{2}\right) }{2}\left\vert \psi
\left( \mathbf{r}\right) \right\vert ^{2}\right.  \notag \\
&-&\left.{\frac{\omega_0^2 r^2}{2}}\left\vert \psi \left( \mathbf{r}\right)
\right\vert ^{2}+ \frac{\alpha }{2}\left\vert \psi \left( \mathbf{r}\right)
\right\vert ^{2}\int \frac{\left\vert \psi \left( \mathbf{r}^{\prime
}\right) \right\vert ^{2}d\mathbf{r}^{\prime }}{1+\left\vert \mathbf{r}-%
\mathbf{r}^{\prime }\right\vert ^{6}}\right] d\mathbf{r}  \label{free energy}
\end{eqnarray}%
where $L_{z}=-i(x\partial _{y}-y\partial _{x})$ is the $z$-component angular
momentum operator and $F_0=F(\omega_{0}=0)$ is the free energy of the system
without rotation. When $\omega _{0}\ll 1$, $F(\omega_{0})$ can be expanded
as $F(\omega _{0})=F_{0}(\psi _{g})-I\omega _{0}^{2}/2$ with $\psi _{g}$,
taken to be real, being the ground state of $F_{0}$. 
Since classical moment of inertia is given by $I_{0}=\int \psi _{g}^{2}r^{2}d%
\mathbf{r}$, we obtain for $\omega _{0}\ll 1$, 
\begin{equation}
\text{NCRIF}=\frac{\int \left[ \left\vert \left( \nabla -i\omega _{0}\mathbf{%
e}_{z}\times \mathbf{r}\right) \bar{\psi} \right\vert ^{2}-\left( \nabla
\psi _{g}\right) ^{2}\right] d\mathbf{r}}{\omega _{0}^{2}\int \psi
_{g}^{2}r^{2}d\mathbf{r}},  \label{NCRIF}
\end{equation}%
where $\bar{\psi}$ is the ground state of $F(\omega_{0})$. In arriving (\ref%
{NCRIF}), we have assumed $|\bar{\psi}|\simeq \psi_{g}$ for $\omega _{0}\ll
1 $. Therefore NCRIF can be obtained by computing Eq.~(\ref{NCRIF}) with the
solved $\bar{\psi}$ and $\psi_{g}$. Fig.~\ref{fig1}(f) plots the calculated
NCRIF as a function of the strength $\alpha $. It is evident that onset of
crystallization of the BEC droplets is characterized by the drastic drop of
NCRIF, occurring at $\alpha \sim 2300$. No similar drop appears for the
onset of the ring-like supersolid phase occurring at smaller $\alpha$ as a
result of full rotational symmetry.

\begin{figure}[t]
\begin{center}
\includegraphics[width=3.5in]{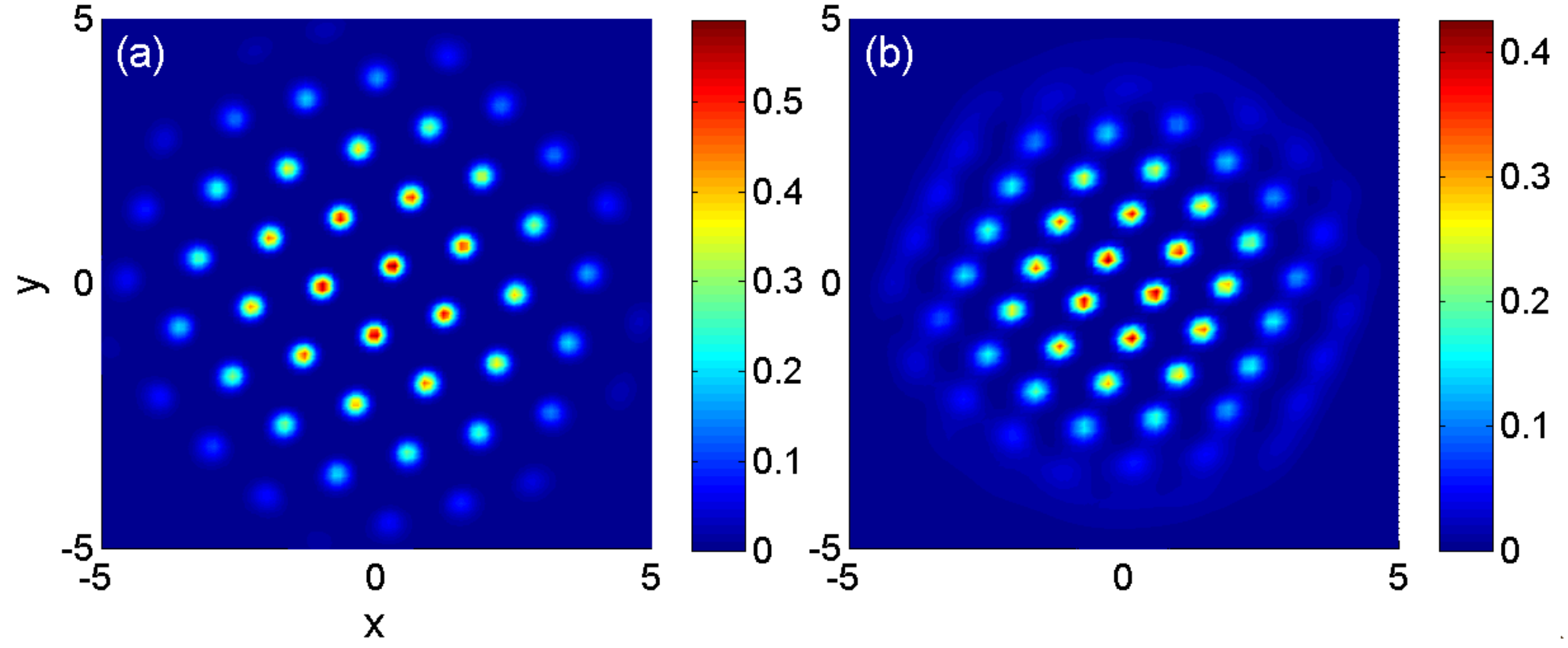}
\end{center}
\par
\vspace{-0.5cm}
\caption{(Color online) Comparison of a hexagonal (a) and a nearly square
(b) lattice structures. Anisotropic ratio $\protect\kappa $ of the
interaction is $1$ ($1.4$) for (a) [(b)] (see text).}
\label{fig3}
\end{figure}

By relaxing the originally frozen $z$ direction potential and/or increasing
the particle number, density starts to modulate in $z$ direction and
eventually forms a multilayer structure. Figs.~\ref{fig2}(a) and \ref{fig2}%
(b) compare formations of a monolayer and a bilayer lattice. The inset in
Fig.~\ref{fig2}(b) indicates clearly that such a bilayer structure is an AB
stack.

It is also interesting to note that when Rydberg dressing interaction
becomes anisotropic, hexagonal lattice can shift to a nearly square lattice
due to distortion of the interaction. This can occur when an external static
electric field $\mathbf{E}$ is applied to the system (Stark effect) to which
two-photon mechanism will acquire an anisotropic component for the
interaction, as compared to the purely isotropic case with $\mathbf{E}=0$ 
\cite{zoller}. Fig.~\ref{fig3} shows a nearly (though not perfectly) square
lattice by using an interaction of the Heaviside-function form $\sim\theta (
1-\sqrt{x^{2}+\kappa ^{2}y^{2}+z^{2}})$ instead of $\sim1/(r^{6}+1)$ as in
Eq.~(\ref{dimesionless}). Here $\kappa$ corresponds to the anisotropic
ratio. The static electric field is considered applied along the $y$ axis.
Without losing the generality, while Heaviside-function interaction make the
simulation of anisotropy more conveniently, it does capture a roton minimum
in the excitation spectrum.

\begin{figure}[t]
\begin{center}
\includegraphics[height=3.0485in,width=3.1401in]{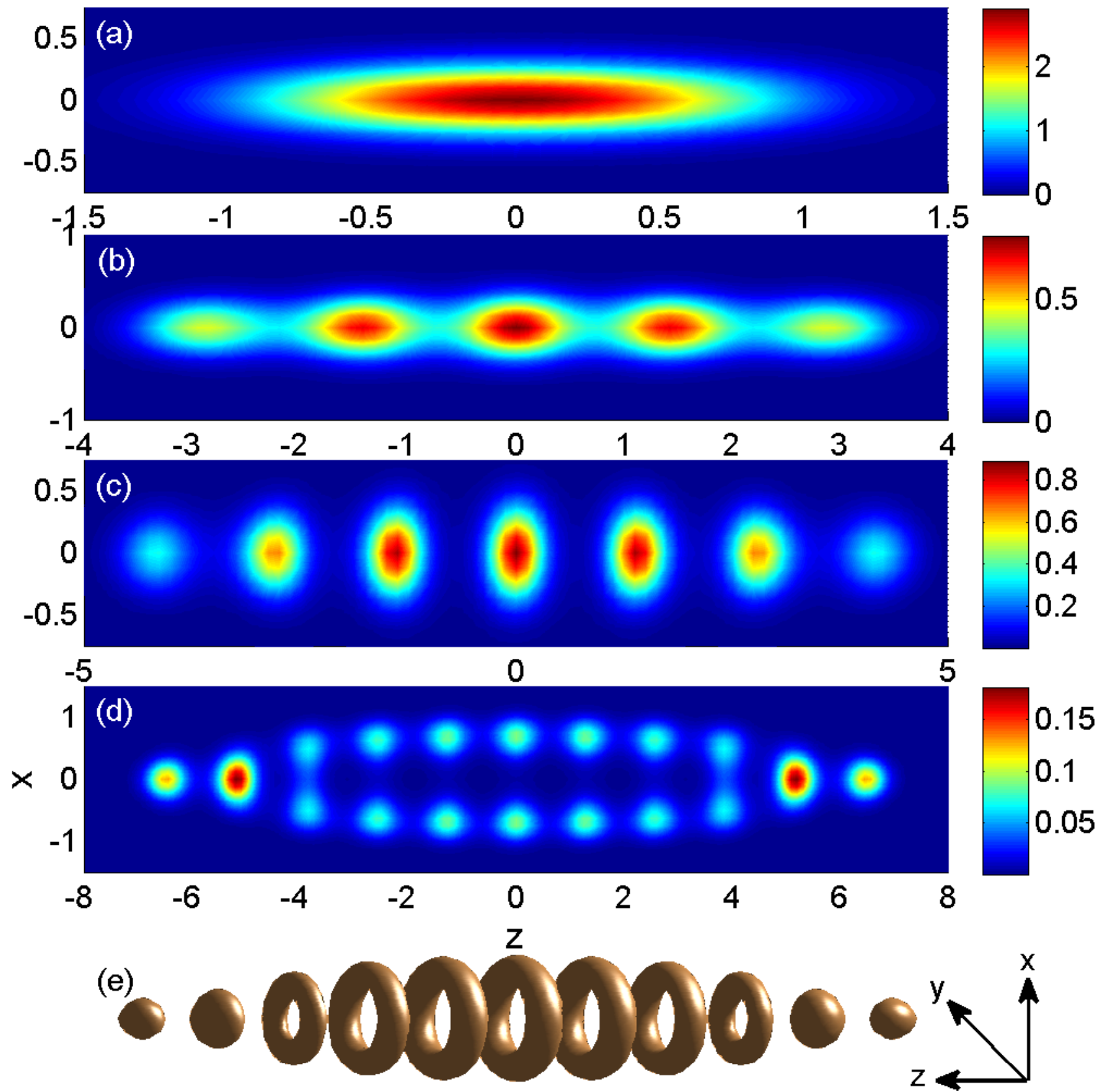}
\end{center}
\par
\vspace{-0.5cm}
\caption{(Color online) Density modulations of a quasi-1D condensate with $%
\protect\omega =15$, $\protect\lambda =0.2$, and (a) $\protect\alpha =10$,
(b) $250$, (c) $500$, and (d) $2000$. (e) is the isosurface of modulation
(d) with isovalue $0.05$.}
\label{fig4}
\end{figure}

The condensate ground state in a quasi-1D system is also investigated with $%
\omega =15$ and $\lambda =0.2$. Fig.~\ref{fig4} shows the modulation of
condensate density of a quasi-1D system by varying the strength of
dressing-induced interaction $\alpha $. When $\alpha $ is small, the system
displays superfluidity, and the density has a central peak [see Fig.~\ref%
{fig4}(a)]. As $\alpha $ increases, the density starts to modulate and has
multiple peaks. Fig.~\ref{fig4}(b) shows that there are five peaks in the
condensate, along the axial direction. With a sufficiently large $\alpha $,
the condensate spontaneously crystallizes in the axial direction [see Fig.~%
\ref{fig4}(c)]. As $\alpha $ increases further, the condensate starts to
modulate in the frozen direction, forming a crystalline structure with a
central hole [see Figs.~\ref{fig4}(d) and \ref{fig4}(e)]. If $\alpha $ is
extremely large, the condensate starts to cluster in the originally frozen
direction, and these clusters form a gyroidal chain.

In summary, based on the Gross-Pitaevskii treatment, spontaneously
crystalline ground states, called quantum crystals, are numerically studied
for a trapped Rydberg-dressed Bose-Einstein condensate. In a quasi-2D
system, a hexagonal droplet lattice characterized by a drastic drop of the
non-classical rotational inertia is shown when dressing interaction is
sufficiently large. By relaxing the originally frozen axis, an AB stacking
bilayer lattice is observed. We also show that by applying a static electric
field to make the interaction anisotropic, a nearly square droplet lattice
can be obtained.

We are grateful to Yu-Ching Tsai for many helpful discussions. This work was
supported by National Science Council of Taiwan (Grant No.
98-2112-M-018-001-MY2). We also acknowledge the support from the National
Center for Theoretical Sciences, Taiwan.


\begin{thebibliography}{99}
\bibitem{Uk2D} $\tilde{U}_{\mathrm{2D}}(k_{\rho })$ is Fourier transform of $%
U_{\mathrm{2D}}(\vec{\rho}-\vec{\rho}^\prime)\equiv \int \varphi
_{g}(z)^{2}\varphi _{g}(z^{\prime })^{2}/[1+|\mathbf{r}-\mathbf{r}^{\prime
}|^{6}]dzdz^{\prime }$. As the softened range of $1/[1+|\mathbf{r}-\mathbf{r}%
^{\prime }|^{6}]$ is much longer than the width of $\varphi _{g}^{2}$, i.e., 
$\omega \lambda \gg 1$, the integration is well approximated by $U_{\mathrm{%
2D}}(\vec{\rho}-\vec{\rho}^\prime)\approx 1/[1+|\vec{\rho}-\vec{\rho}%
^\prime|^{6}]\times \int \varphi _{g}(z)^{2}\varphi _{g}(z^{\prime
})^{2}dzdz^{\prime }= 1/[1+|\vec{\rho}-\vec{\rho}^\prime|^{6}]$.
Consequently $\tilde{U}_{\mathrm{2D}}(k_{\rho })=(\pi/3)G_{06}^{40}({%
k_{\rho}^{6}}/{6^{6}} \QATOPD\vert.{ {-}}{0,1/3,2/3,2/3,0,1/3})$ with $%
G_{pq}^{mn}(z\QATOPD%
\vert.{a_{1},...,a_{n},a_{n+1},...,a_{p}}{b_{1},...,b_{m},b_{m+1},...,b_{q}}) 
$ the Meijer G function.
\end{thebibliography}

\end{document}